\documentclass{article}
\usepackage{spconf,amsmath,graphicx}
\usepackage{mathrsfs}
\usepackage{amsfonts}
\usepackage{bm}
\usepackage{enumitem}
\usepackage{url}
\newcommand{\zchd}{z_{\text{chd}}}
\newcommand{\zatxt}{z_{\text{txt}}^{\text{aud}}}
\newcommand{\zstxt}{z_{\text{txt}}^{\text{sym}}}

\title{Audio-to-symbolic Arrangement via Cross-modal Music Representation Learning}

\name{Ziyu Wang$^{12}$ \quad Dejing Xu$^{2}$ \quad Gus Xia$^{1}$ \quad Ying Shan$^{2}$}
\address{$^{1}$Music X Lab, NYU Shanghai \quad $^{2}$ARC Lab, Tencent PCG}

\begin{document}
\maketitle
\begin{abstract}
Could we automatically derive the score of a piano accompaniment based on the audio of a pop song? This is the audio-to-symbolic \textit{arrangement} problem we tackle in this paper. A good arrangement model should not only consider the audio content but also have prior knowledge of piano composition (so that the generation ``sounds like'' the audio and meanwhile maintains musicality). To this end, we contribute a cross-modal representation-learning model, which 1) extracts chord and melodic information from the audio, and 2) learns texture representation from both audio and a \textit{corrupted} ground truth arrangement. We further introduce a tailored training strategy that gradually shifts the source of texture information from corrupted score to audio. In the end, the score-based texture posterior is reduced to a standard normal distribution, and only audio is needed for inference. Experiments show that our model captures major audio information and outperforms baselines in generation quality. \footnote{Code and demos can be accessed via \url{https://github.com/ZZWaang/audio2midi}.}

\end{abstract}
\begin{keywords}
cross-modal representation, automatic arrangement, disentangled representation
\end{keywords}

\section{Introduction}
\label{sec:intro}

Piano arrangement is widely used in practice to reproduce various complex music signals. The key idea is to transform the original music, usually represented by an audio mixture or full score of a band, into a piano score (that can be performed using only 2 or 4 hands) without loss of major music information. For example, piano reductions are made for classical orchestral music and piano covers are created for pop songs. 
A good arrangement is not merely a transcription of the original audio mapped onto the keyboard, but also a realization of the original content that makes musical sense as a composition in the new instrument.

In the paper, our focus is automatic \textit{audio-to-symbolic} arrangement. Unlike automatic music transcription \cite{benetos2013automatic, benetos2018automatic}, the task aims at the symbolic generation according to the audio content which consists of an arbitrary set of instruments and may contain timbral effects that cannot be easily transcribed. Existing systems mainly rely on rule-based or simple statistical models \cite{percival2015song2quartet, takamori2017automatic, takamori2019audio, ariga2017song2guitar}, where audio-analysis and symbolic-generation modules are more or less independent and loosely connected by some bottleneck audio features. Such design often suffers from two limitations. First, the arrangement patterns are usually very rigid since both the bottleneck audio features and the accompaniment patterns are largely pre-defined. Second, nuanced features such as groove patterns and bass lines are difficult to be 
modeled using existing MIR techniques.

We consider end-to-end audio-to-symbolic deep generative modeling a better choice, as neural-based modeling potentially enables more flexible symbolic generation and an end-to-end architecture allows nuanced features to flow from audio to symbolic modal. On the other hand, we are faced with a great challenge --- the relation between audio and its possible symbolic arrangement is essentially \textit{one-to-many} and the prior knowledge of a good piano composition is only partially present in the audio. In other words, a naive supervised-learning model would easily get confused by the noisy audio-symbolic pairs and collapse to certain specific accompaniment patterns. %

To solve the problem, we propose a cross-modal representation learning framework, in which the input is the audio of a pop-song accompaniment under arbitrary instrumentation, while the output is an arrangement in MIDI format. The model encodes a cross-modal representation from both audio and symbolic modals and decodes the information back to the symbolic domain. The latent representation contains the audio content and reflects prior knowledge about symbolic composition. During pre-training, we initialize the model to be an almost pure symbolic-to-symbolic variational autoencoder. By gradually corrupting the input and strengthening the variational constraints, the model is trained to lean more towards the audio. During the fine-tuning, we could provide the symbolic side with either Gaussian noise or information from previous bars to make the model fully dependent on audio or being autoregressive. 

\begin{figure*}[h!]
\vspace{0.3cm}
\begin{minipage}[b]{0.575\linewidth}
  \centering
  \centerline{\includegraphics[width=0.98\linewidth]{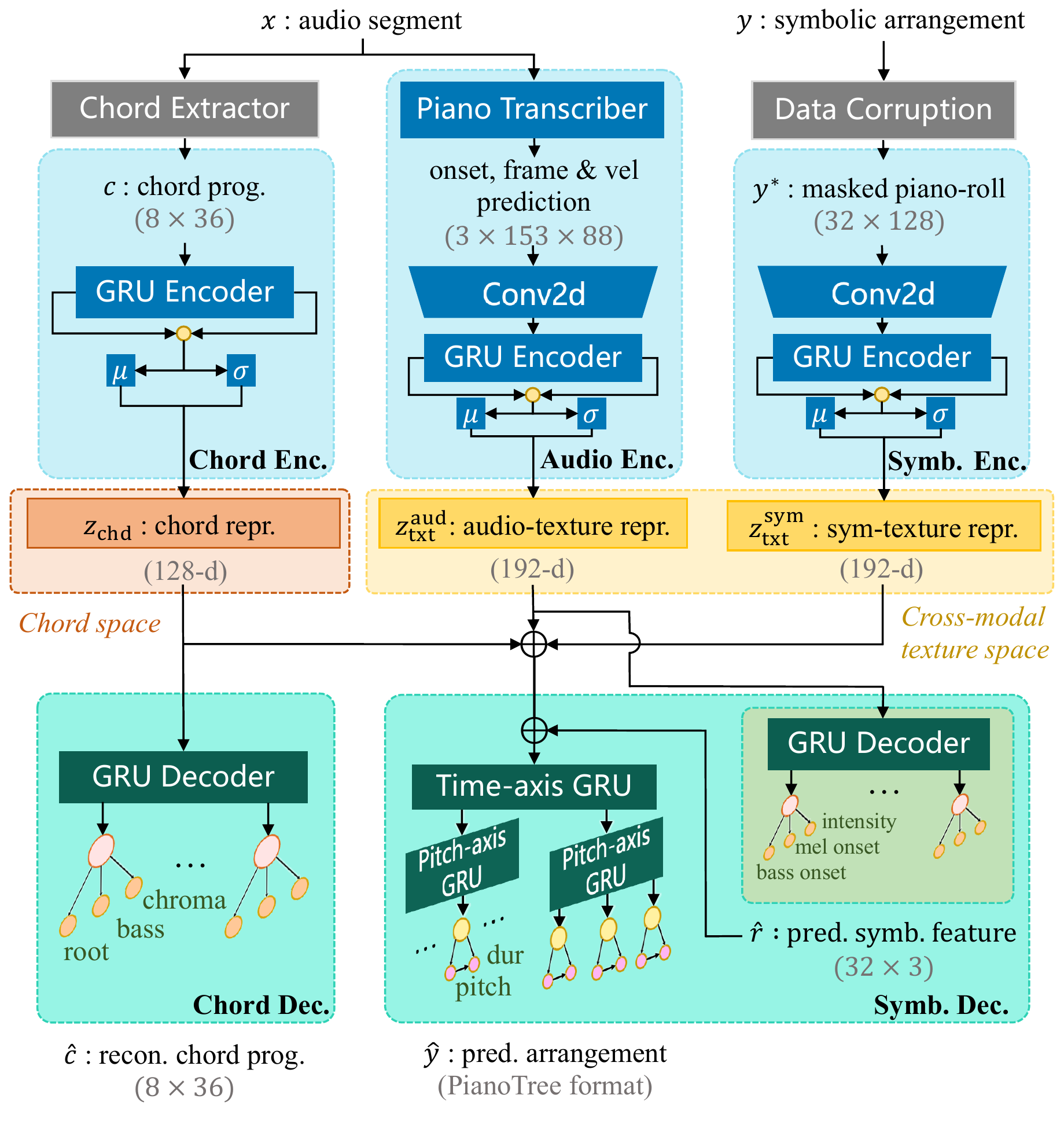}}

  \centerline{(a) Model architecture.}\medskip
\end{minipage}
\hfill
\hspace{-0.13cm}
\begin{minipage}[b]{0.35\linewidth}
  \centering
  \centerline{\includegraphics[width=1\linewidth]{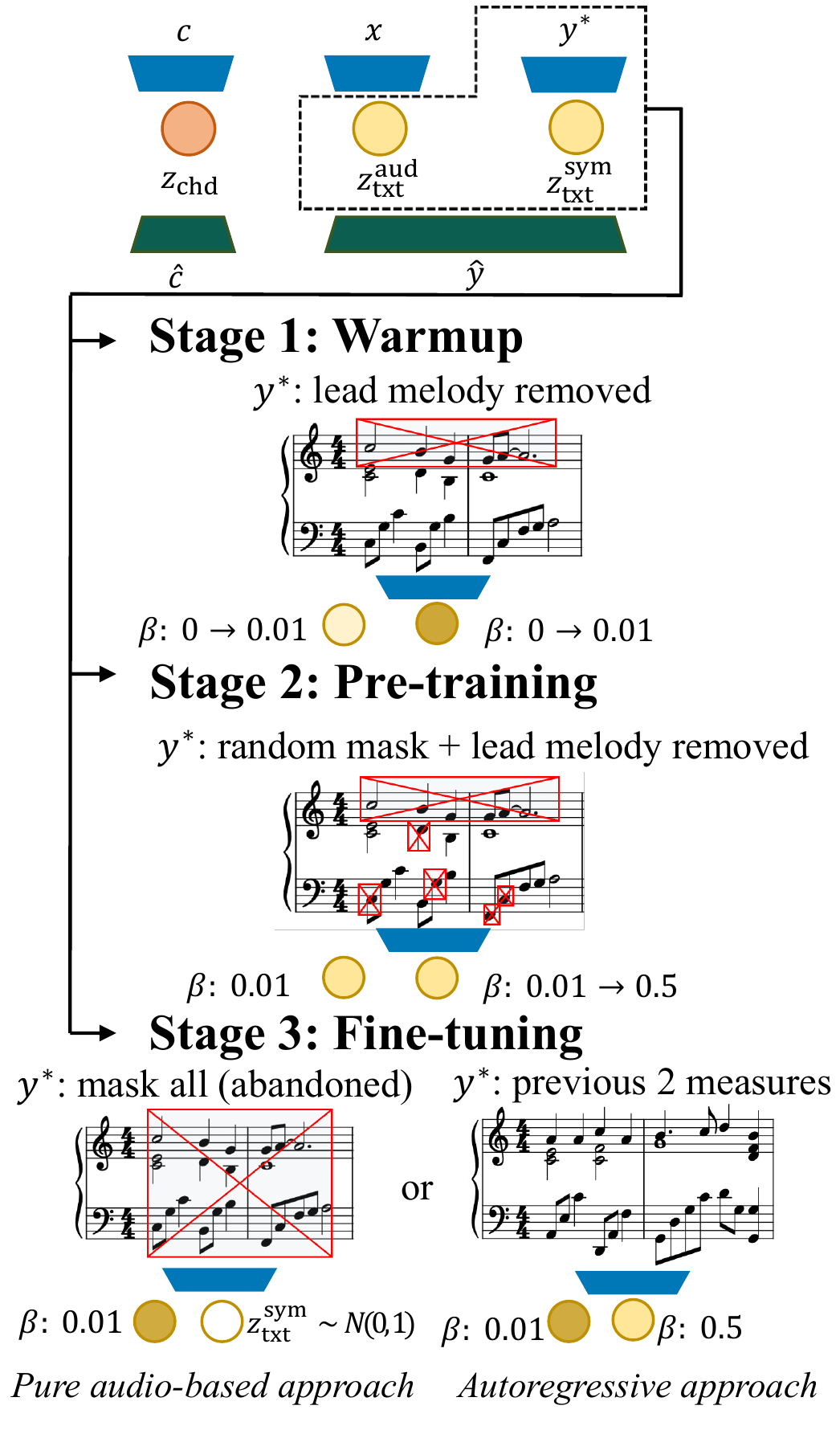}}
  \centerline{(b) Training strategy.}\medskip
\end{minipage}

\caption{The proposed model architecture and training strategy.}
\label{fig:diag}
\end{figure*}

In sum, we contribute the first end-to-end approach for automatic audio-to-symbolic arrangement. The quality of the generated samples is significantly higher than baselines and even rated higher than human compositions in terms of creativity. Moreover, the arrangement problem is tackled by a tailored training strategy that optimizes the supervised objective under an unsupervised cross-modal representation learning framework, which can be potentially generalized to other one-to-many supervised training tasks.

\section{RELATED WORK}
\label{sec:relatedwork}

Automatic arrangement tasks can be conducted based on either symbolic or audio sources. Pure symbolic arrangement is commonly studied using deep generative models and has achieved considerable progress \cite{polydis, simon2018learning, briot2017deep, ke}. In contrast, \textit{audio-to-symbolic} arrangement, which is more related to this paper, is still underresearched. Existing methods are mostly rule-based or rely on hand-crafted statistics. E.g., Takamori et al. extract chords and melodies from audio and pre-define several accompaniment textures \cite{takamori2017automatic, takamori2019audio}. Song2Quartet \cite{percival2015song2quartet} and Song2Guitar\cite{ariga2017song2guitar} further introduce matching probabilities between audio and score and use dynamic programming to search the notes. These models often lead to rigid patterns and the musicality cannot yet serve for practical purposes.

An alternative shortcut to achieve arrangement is timbre style transfer \cite{engel2020ddsp, wang2019performancenet, hung2019musical, lin2021unified, dai2018music}, in which a latent timbre space is first learned and then transferred to piano timbre during inference. However, existing models are either constrained to monophonic instruments or based on synthetic audio data. Real-world audio is more complicated in instrumentation and the integration of audio content with piano composition techniques is still an open problem.

\section{Method}
\label{sec:method}

We aim to learn \textit{chord representation}, \textit{audio-texture representation} from audio and \textit{symbolic-texture representation} from its paired symbolic arrangement. 
In this paper,  we consider 2-bar audio segments (provided with beat annotation) and symbolic arrangement in $\frac{4}{4}$ time signature. The symbolic arrangement is represented under $\frac{1}{4}$ beat resolution.

\subsection{Model Architecture}
\label{subsec: architecture}
Figure~\ref{fig:diag}(a) shows the overall model architecture which adopts an encoder-decoder architecture and contains five parts: 1) a chord encoder, 2) a chord decoder, 3) an audio encoder, 4) a symbolic encoder, and 5) a symbolic decoder. We consider our model a cross-modal extension of a symbolic-domain chord and texture disentanglement study \cite{polydis}.

The chord encoder adopts a GRU layer to encode 128-d chord representation $\zchd{}$ from a chord progression, which is extracted from the audio using an existing chord extraction algorithm \cite{JiangChordExtractor}. The chord decoder which reconstructs the input chord progression is introduced as a 
mechanism to avoid \textit{posterior collapse} of $\zchd{}$.   

The input to the audio encoder is an 8-beat long audio segment, time-stretched and resampled to 95 BPM with a sample rate of 16kHz. We first use a piano transcriber architecture \cite{hawthorne2017onsets} to embed the audio feature into a stack of piano-roll-like matrices of onset, frame and velocity predictions. Then, we use a 2D convolution layer followed by a GRU layer to encode 192-d audio-texture representation $\zatxt{}$. The same model structure is applied to the symbolic encoder to extract 192-d symbolic-texture representation $\zstxt{}$ from a corrupted ground truth piano-roll. The data corruption is controlled by a tailored training strategy introduced in section~\ref{subsec:curriculum}.

The symbolic decoder takes in the concatenation of $\zchd{}$, $\zatxt{}$ and $\zstxt{}$ and decodes the symbolic arrangement in a hierarchical manner using the decoder module of PianoTree VAE \cite{pianotree}, the state-of-the-art polyphonic representation learning model. Besides the latent codes, the PianoTree decoder also takes in a time series of symbolic features, which is predicted from $\zatxt{}$ \textit{only} to enhance audio information retrieval. Specifically, we explicitly constrain $\zatxt{}$ to predict three symbolic features: \textit{bass onset}, \textit{melody onset}, and \textit{rhythmic intensity}, which usually strongly correlate with audio rhythmic information of bass drum, lead melody and groove patterns, respectively. Both bass onset and melody onset are time series of onset probabilities, and rhythmic intensity is a time series of scalar values. The predicted features are fed to the corresponding time steps of the time-axis GRU in the PianoTree decoder. Similar method is also used to achieve disentanglement in \cite{ec2vae}.

\subsection{Training Objective}
The loss terms in our model include 1) reconstruction losses of chord, arrangement, and symbolic features, and 2) KL losses between all three latent factors with standard normal distributions. Our model is essentially a \textit{conditional variational autoencder}, since the loss function can be formalized as the \textit{evidence lower bound} (ELBO) of the conditional probability $p(y|x)$, where $x$ is the audio and $y$ is the arrangement.

The posterior distribution of the conditional VAE is defined as the product of the three encoder models:
\begin{align*}
    q_{\boldsymbol\phi}(\mathbf{z}|x, y) := q_{\phi_1}(\zchd{}|c)q_{\phi_2}(\zatxt{}|x)q_{\phi_3}(\zstxt{}|y),
\end{align*}
where $\mathbf{z}:= [\zchd{}, \zatxt{}, \zstxt{}]$, and $\boldsymbol\phi := [\phi_1, \phi_2, \phi_3]$ denotes the encoder parameters. Note that the chord progression $c$ is a deterministic transform from $x$ and is therefore absent in $q_{\boldsymbol\phi}(\mathbf{z}|x, y)$. The reconstruction distribution is defined as the product of the three reconstruction terms:
\begin{align*}
    p_{\boldsymbol\theta}(x|\mathbf{z}) := p_{\theta_1}(c|\zchd{})p_{\theta_2}(y|\mathbf{z}, r)p_{\theta_3}(r|\zatxt{}),
\end{align*}
where $\boldsymbol\theta := [\theta_1, \theta_2, \theta_3]$ denotes the decoder parameters, $r$ denotes the ground truth symbolic features, and $p_{\theta_1}(c|\zchd{})$ is interpreted as a regularizer to the output distribution. Finally, the loss function is:
\begin{align*}
    \mathcal{L}_\beta(\boldsymbol\theta, \boldsymbol\phi; x) = &-\mathbb{E}_{\mathbf{z} \sim q_{\boldsymbol\phi}(\mathbf{z}|x, y)}\Bigl[\log p_{\boldsymbol\theta}(x|\mathbf{z})\Bigr] \\  
    & + \beta \text{ KL}(q_{\boldsymbol\phi}(\mathbf{z}|x, y)||p(\mathbf{z})),
\end{align*}
where $p(\mathbf{z})$ is a 512-d standard normal prior and $\beta$ is the KL annealing parameter.

\subsection{Training Strategy}
\label{subsec:curriculum}

We propose a training strategy (shown in Figure~\ref{fig:diag}(b)) to balance information from the audio encoder and the symbolic encoder so that the training starts with the unsupervised symbolic reconstruction task and shifts to the supervised audio-to-symbolic task. Particularly, there are three stages:

\begin{description}[style=unboxed,leftmargin=0cm, listparindent=\parindent, parsep=0pt]
\item[Stage 1, Warm-up:] The lead voice of the ground truth arrangement is masked, and $\beta$ increases from 0 to 0.01 for all three latent factors. The model is therefore enforced to predict melody solely from the audio.
\item[Stage 2, Pre-training:] Besides melody, the rest of the notes are randomly masked under probability ranging from 0.5-0.8, where lower pitches have a higher probability to be masked. Meanwhile, $\beta$ increases from 0.01 to 0.5 for $\zstxt{}$ and keeps $0.01$ for the other two factors. The model is expected to learn more information from the audio.
\item[Stage 3, Fine-tuning:] The model can be purely audio-dependent at this stage: we completely abandon the symbolic encoder by sampling $\zstxt{}$ from a standard normal distribution. Alternatively, we can also feed the arrangement of the previous two measures to the symbolic encoder to make the model autoregressive.
\end{description}

\section{Experiments}
\label{sec:experiments}

\begin{figure*}[htb]
\vspace{0.3cm}
\centering
 \centerline{
 \includegraphics[width=1\linewidth]{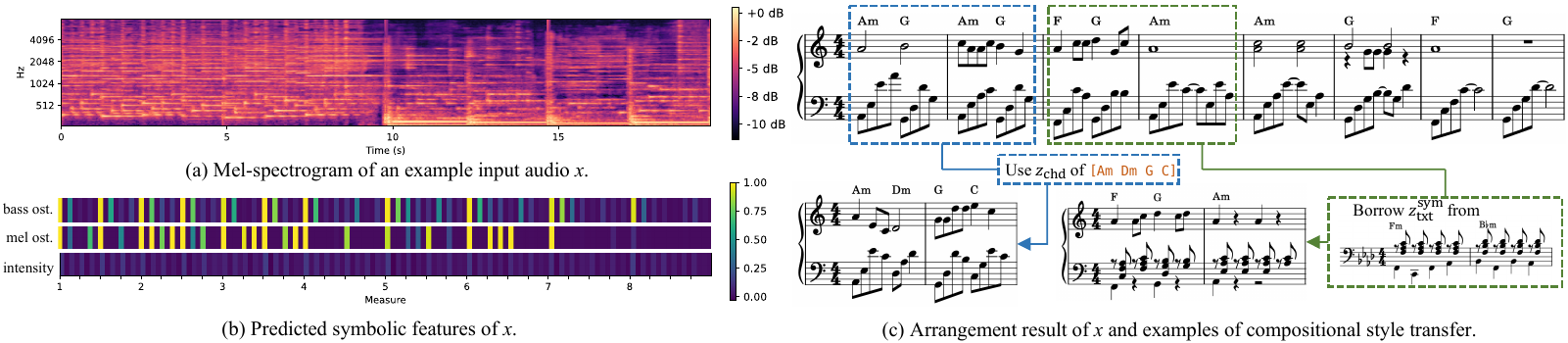}}

 \caption{An example of automatic arrangement based on the audio of an 8-bar excerpt from \textit{1001 Nights} by Samuel Tai. }
 \label{fig:example}
\end{figure*}

\subsection{Implementation Detail}
We train our model on the POP909 dataset \cite{pop909}, which contains about 1K MIDI files of pop song arrangements with time-aligned audios. We use the piano accompaniment MIDI tracks and keep the pieces with $\frac{2}{4}$ and $\frac{4}{4}$ meters and cut them into 8-beat music segments. The audio is also sliced into 8-beat segments and the vocal is removed by the Spleeter source separation algorithm \cite{spleeter2020}. In all, we have 66K samples. We randomly split the dataset (at song level) into training set (90\%) and test set (10\%). All training samples are further augmented by transposing to all 12 keys. The chord, beat, and melody track annotations are all included in the dataset while the ground truth bass onset is defined to be the occurrence of MIDI pitch lower than 48, and rhythmic intensity is the number of simultaneous onsets normalized by a constant.

The piano transcriber used in our model is pre-trained on the MAESTRO dataset \cite{hawthorne2018enabling}. The model contains 62M trainable parameters in total, including 26M parameters in the piano transcriber. We use a batch size of 64 and Adam optimizer \cite{adamopt} with a scheduled learning rate from 4e-4 to 6e-4.

\subsection{Arrangement Example}
A 16-bar arrangement example (by predicting every 2 bars independently) is shown in Figure~\ref{fig:example}. The audio has a lead instrument and frequent bass note changes at the beginning, and a less intense groove halfway to the end (Figure~\ref{fig:example}(a)). These features are captured in the symbolic feature prediction (Figure~\ref{fig:example}(b)), as well as the symbolic arrangement (Figure~\ref{fig:example}(c)), where we see melody with arpeggio texture in mm.~1-4, and arpeggio with decreasing intensity in mm.~5-8.

We also demonstrate the model (after the pre-training stage) is capable of the \textit{compositional style transfer} tasks \cite{dai2018music} via replacement of the disentangled factors \cite{polydis}. First, we change the chords in mm.~1-2 to another chord progression [Am, Dm, G, C] represented by $\zchd{}$ (indicated by the blue arrow), and the generation changes to the desired progression while maintaining the original texture. Then, we replace mm.~3-4 with a new symbolic texture represented by $\zstxt{}$ (indicated by the green arrow), and the left-hand texture changes correspondingly while the harmony and the right-hand melody contour are kept unchanged.

\subsection{Subjective Evaluation}
We compare our proposed method with three baselines. The first two baselines adopt the common supervised approach, implemented with only the audio encoder and the PianoTree decoder with or without KL loss. The third baseline is solely chord-dependent, by setting $\zatxt{}$ to zero. 

We invite people to subjectively rate the generation quality through a double-blind online survey. During the survey, the subjects listen to 6 groups of samples. In each group, the original audio is played, followed by the generated samples and the ground truth composition in random order. Both the order of groups and the sample order within each group are randomized. After listening to each sample, the subjects rate them based on a 5-point scale from 1 (very low) to 5 (very high) according to four criteria: \textit{faithfulness} (to the original audio), \textit{creativity}, \textit{naturalness} and overall \textit{musicality}.

\begin{figure}[t!]
 \centerline{
 \includegraphics[width=0.9\columnwidth]{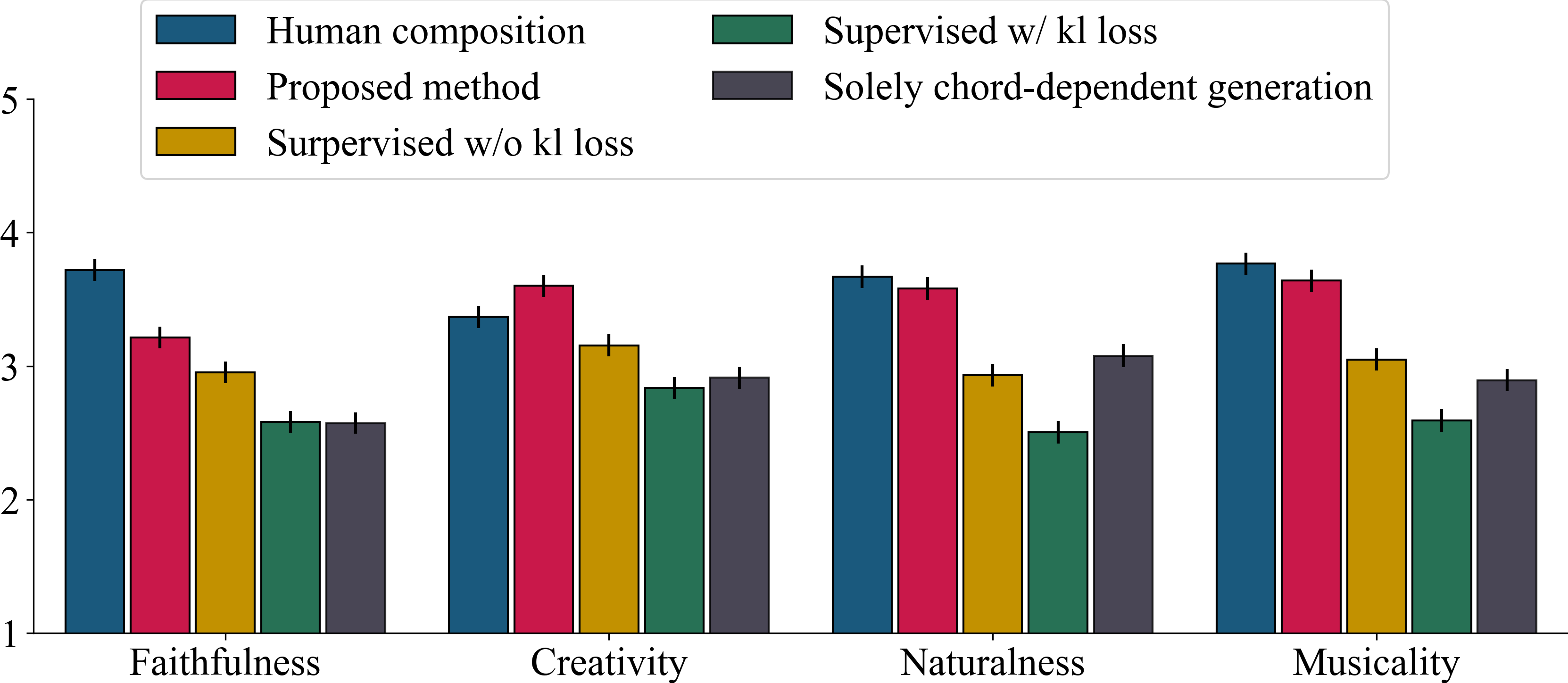}}
 \caption{Subjective evaluation results.}
 \label{fig:barplot}
\end{figure}

A total of 26 subjects (8 females and 18 males) with different musical backgrounds have completed the survey. Figure~\ref{fig:barplot} shows the result where the heights of the bars represent the means of the ratings and the error bars represent the confidence interval computed via within-subject ANOVA. The result shows that the proposed model is significantly better than the baseline models in terms of all four criteria, and the creativity is even significantly better than human composition (with p-value $< 0.005$).

\section{Conclusion}
\label{sec:conclusion}
We have contributed a cross-modal representation learning framework as the first end-to-end approach to accomplish the audio-to-symbolic automatic arrangement problem. Experimental results show that our model is able to capture harmonies, melody lines, and groove patterns from the audio without loss of musicality. The main novelty lies in the cross-modal training strategy that gradually shifts the input source from one modal to the other. We see such kind of tailored self-supervision control as a bridge between unsupervised learning tasks and supervised training. In the future, we seek more flexible cross-modal methods and make the audio-to-symbolic conversion more controllable.

\bibliographystyle{IEEEbib}
\bibliography{strings,refs}

\end{document}